\documentstyle[aaspptwo,dec]{article}

\def\al3{$\alpha_{3}$}

\begin{document}

\twocolumn

\title{A New Method for Obtaining Binary Pulsar Distances and its
Implications for Tests of General Relativity}
\author{J. F. Bell} \affil{Mount Stromlo and Siding Spring Observatories,
Institute of Advanced Studies,\\ Australian National University, Private
Bag, Weston Creek, ACT 2611, Australia \\email: bell@mso.anu.edu.au}
\author{M. Bailes} 
\affil{Australia Telescope National Facility, 
CSIRO, PO Box 76, Epping, NSW 2121,
Australia\\email: mbailes@atnf.csiro.au} 

\begin{abstract}
We demonstrate how measuring orbital period derivatives can lead to more
accurate distance estimates and transverse velocities for some nearby binary
pulsars.  In many cases this method will estimate distances more accurately
than is possible by annual parallax, as the relative error decreases as
${t^{-5/2}}$. Unfortunately, distance uncertainties limit the degree to
which nearby relativistic binary pulsars can be used for testing the general
relativistic prediction of orbital period decay to a few percent.
Nevertheless, the measured orbital period derivative of PSR B1534+12 agrees
within the observational uncertainties with that predicted by general
relativity if the proper-motion contribution is accounted for.
\end{abstract}


\section{Introduction}
\label{s:accel}

The most common method for determining the distances to radio pulsars is
based on their dispersion measure and models of the Galactic distribution of
free electrons (\cite{tc93}). These distance estimates typically have an
uncertainty of 30\%. Distances may also be determined by measuring annual
parallax, based on either timing (\cite{rt91a}) or interferometric
measurements (\cite{gtwr86}). H{\footnotesize I} absorption by interstellar
hydrogen is also a common distance indicator (\cite{fw90}).  However, no pulsar
has a distance estimate more accurate than $\sim10$\%, and for all but two,
the errors are greater than 20\%.

Any acceleration of a pulsar along the line of sight will change the
observed pulse period derivative $\dot P$. As Shklovskii (1970) pointed out,
an apparent acceleration occurs when the proper motion is significant. The
magnitude of this contribution is $\dot P_{(pm)}/P = v^2/(cd)$, where $P$ is
the pulse period, $v$ is the transverse velocity, $d$ is the pulsar
distance, and $c$ is the speed of light. For many millisecond pulsars, this
effect is of similar magnitude to the intrinsic pulse period derivative,
making it hard to determine accurately from timing data either the intrinsic
pulse period derivative (\cite{ctk94}), or the distance and transverse
velocity.

\section{Distances and Velocities}

This apparent acceleration also applies to orbital period derivatives, and
the contribution is $\dot{P}_{b(pm)} / P_{b} = v^2 / (cd)$, where $P_{b}$ is
the orbital period (\cite{dt91}). In fact, for many nearby millisecond
pulsars, it is expected to completely dominate future observed orbital
period derivatives. This means that $v^2/d$ can be obtained and, when
combined with the measured proper motion $\mu = v/d$, the distance and
transverse velocity can be easily separated. Hence, the proper-motion
contribution to the pulse period derivative can also be determined, giving
accurate estimates of the intrinsic pulse period derivative and hence the
magnetic field strengths, ages and spin-down luminosities of binary
millisecond pulsars.

The amplitude and functional form of the residuals from a least-squares fit
to the observed pulse arrival times, if one parameter is set to zero, is
often called the ``timing signal'' for that parameter.  For proper motion,
the timing signal is often relatively large, with its amplitude increasing
linearly with time. With continued measurement therefore, its relative error
decreases as $t^{-1.5}$.  The peak-to-peak amplitude $\Delta T_{(pm)}$ of
the timing signal owing to the contribution of the proper motion to the
observed orbital period derivative, is
\begin{equation}
{\Delta T_{pm}} = {a \sin i \over c} {2\pi\over P_b} {v^2\over cd} t^2
\end{equation}
where $a$ is the semi-major axis of the pulsar's orbit and $i$ is the
orbital inclination. The accuracy of distances obtained in this way are
limited by the accuracy of the orbital period derivative measurements. Their
accuracy, and therefore the accuracy of distances improve as $t^{-2.5}$. The
fact that the relative error in both of these critical parameters decreases
in such a spectacular fashion with time demonstrates the power of this
method for determining distance and transverse velocity.

Table \ref{t:pbdot} shows the predicted size of the timing signal $\Delta
T_{pm}$ after 10 years of regular timing observations for a selection of
binary millisecond pulsars. Where the proper motion was not available, we
used the median transverse velocity for millisecond pulsars of 69
km~s$^{-1}$ was used.  Also shown is the timing signal due to parallax
(\cite{rt91a}), ${\Delta T_{\pi}} = r^{2} \cos^{2} \theta /(2cd)$, where $r$ is
the radius of the Earth's orbit and $\theta$ is the angle between the line
of sight to the pulsar and the Earth's orbital plane.  After 10 years, the
new method will provide better distance estimates than parallax
measurements.  This is possible because $\Delta T_{pm} \propto t^{2}$, while
$\Delta T_{\pi}$ is constant. If an rms timing residual of 1.0 $\mu$s could
be obtained it would be possible to determine distances this way for several
of the currently known binary millisecond pulsars.  On average, for the
pulsars listed in Table \ref{t:pbdot}, 2--3 years of precise timing data
have been recorded by various observers. So, to reap the rewards of this
method a further 7--8 years of precise timing will be required.

\section{Other Contributions to Period derivatives}

Many other effects could contribute to an observed orbital period
derivative; for example, changes in the gravitational constant
(\cite{dt91}), tidal effects (\cite{wil93,aft94,as94}), companion mass loss
(\cite{it86}) and accelerations in globular cluster potentials
(\cite{bra87}). These contributions are indistinguishable from the
proper-motion contribution, and so it is important to determine which of
them are significant. Known pulsars possess one of 5 types of companion: a
neutron star, a main sequence star, a white dwarf, a very low mass helium
star, or a planetary system (\cite{wol94}). Fortunately, binary pulsars with
either white dwarf or neutron star companions are very ``clean'', and their
orbital periods not affected by tidal or mass-loss effects (\cite{it86}).
Systems with low-mass companions such as PSR B1957+20 possess large orbital
period derivatives, possibly caused by tidal effects (\cite{as94}).  The
small $a \sin i$ induced by planetary companions in the pulsar orbit makes
it extremely difficult to measure their orbital period derivatives.  The
only significant contributions to the orbital period derivatives in neutron
star and white dwarf systems are those due to acceleration in the Galactic
potential $\dot{P}_{b(kz)}$, Galactic differential rotation
$\dot{P}_{b(dr)}$, proper motion $\dot{P}_{b(pm)}$, and general relativity
$\dot{P}_{b(gr)}$. Table \ref{t:pbdot} lists those contributions showing,
that the proper-motion term will dominate for many of the binary millisecond
pulsars.

For the nearby millisecond pulsar J0437$-$4715, the uncertainty in
$\dot{P}_{b(kz)}$ is approximately 1\% of $\dot{P}_{b(pm)}$. Hence,
measurement of the orbital period derivative will ultimately provide a
distance estimate which is limited in accuracy to about 1\%. If the distance
could be independently estimated with superior accuracy, it would be
possible to determine the acceleration of the binary in the Galactic
gravitational potential and thereby constrain the distribution and
composition of dark matter (\cite{ff94}). However, pulsars such as PSR
J2317+1439 with large z-heights are probably better suited to such an
exercise, because contribution from the Galactic acceleration in such
pulsars is comparable to the contribution from the proper motion. This
emphasises the importance of monitoring known binary millisecond pulsars and
searching for new ones.

\section{Implications for Tests of General Relativity}

The double neutron-star system PSR B1534+12 has been predicted to provide an
even better relativistic laboratory than the binary pulsar
B1913+16 (\cite{arz95}). Unfortunately, the distance to this pulsar is known
only to an accuracy of some 30\% (\cite{tc93}), and therefore the proper
motion contribution to the orbital period derivative is uncertain by a
similar amount. Recent measurements (\cite{arz95}) indicate that the predicted
orbital period derivative due to gravitational wave emission
$\dot{P}_{b(gr)}$ is $-1.924 \times 10^{-13}$, whereas the observed value
$\dot{P}_{b(obs)}$ is only $-1.5 \pm 0.3 \times 10^{-13}$.  Using the
dispersion-measure distance of $0.68 \pm 0.2$ pc and the measured proper
motion, the contribution to the observed value from the proper motion
is $\dot{P}_{b(pm)} =0.40 \pm 0.12 \times 10^{-13}$. Since $\dot{P}_{b(gr)}
= \dot{P}_{b(obs)} - \dot{P}_{b(pm)} = -1.9 \pm 0.3 \times 10^{-13}$, the
observed value is in excellent agreement with the general relativistic
prediction. Unless the distance estimate can be improved, the orbital period
decay due to the emission of gravitational waves cannot be verified to
better than $\sim$ 5\% in the PSR 1534+12 system. This is a surprising
result, which underlines the importance of obtaining independent distance
estimates to this system.

\onecolumn

\begin{table}[htb]
\begin{center}
\begin{tabular}{|l|rr|cccc|}
\hline\hline
\multicolumn{1}{|c|}{Pulsar} & $\Delta T_{pm}$ & $\Delta T_{\pi}$ &
\multicolumn{4}{|c|}{Contributions to $\dot{P}_{b}/P_{b(obs)}$ 
($ \times 10^{-19}$ s$^{-1}$)} \\  
\multicolumn{1}{|c}{Name} & 
\multicolumn{2}{|c|}{($\mu s$)} 
& $\dot{P}_{b(kz)}/P_{b} $ & $\dot{P}_{b(dr)}/P_{b}$ 
& $\dot{P}_{b(gr)}/P_{b}$ & $\dot{P}_{b(pm)}/P_{b}$ \\ \hline
J1713+0747$^1$     & 0.3  & 1.0 & \dec --0.89  & \dec --0.29  &      $\sim$ 0 & \dec 0.88 \\
B1855+09$^2$       & 0.5  & 0.8 & \dec --0.029 & \dec --0.002 & \dec  --0.001 & \dec 0.96 \\
J0613--0200$^3$    & 1.5  & 0.4 & \dec --0.33  & \dec --0.67  & \dec  --0.25  & \dec 2.4  \\
J2317+1439$^4$     & 1.8  & 0.5 & \dec --2.2   & \dec  0.89   & \dec  --0.043 & \dec 2.8  \\ 
J2145--0750$^5$    & 2.2  & 2.4 & \dec --1.3   & \dec  0.033  & \dec  --0.009 & \dec 2.1  \\ 
J0751+1800$^6$     & 2.8  & 0.6 & \dec --0.98  & \dec --0.77  & \dec --21.8   & \dec 2.6  \\
J0034--0534$^5$    & 3.4  & 1.2 & \dec --2.8   & \dec  0.17   & \dec  --0.15  & \dec 5.3  \\ 
B1913+16$^7$       & 3.8  & 0.1 & \dec --0.064 & \dec  5.4    & \dec --845.3  & \dec 0.74 \\
J2019+2425$^8$     & 4.5  & 0.7 & \dec --0.11  & \dec  0.45   &      $\sim$ 0 & \dec 12.6 \\ 
J1012+5307$^9$     & 7.0  & 1.5 & \dec --1.7   & \dec  0.16   & \dec  --1.8   & \dec 10.1 \\ 
J1022+02$^{10}$    & 13.5 & 2.0 & \dec --1.8   & \dec  0.048  & \dec  --0.010 & \dec 8.7  \\ 
J1455--3330$^3$    & 22.4 & 1.5 & \dec --0.69  & \dec --0.24  &      $\sim$ 0 & \dec 73.1 \\
J0437--4715$^{11}$ & 28.5 & 1.2 & \dec --0.57  & \dec  0.058  & \dec  --0.004 & \dec 67.5 \\
B0655+64$^2$       & 31.5 & 1.4 & \dec --0.65  & \dec --0.16  & \dec  --2.1   & \dec 10.9 \\
B1534+12$^{12}$    & 70.5 & 1.3 & \dec --1.8   & \dec --0.24  & \dec  --52.8  & \dec 11.0 \\
\hline                                                                  
\end{tabular}                                                           
\end{center}                                                            
\caption[Predicted orbital period derivatives and timing signals]
{Predicted orbital period derivatives and timing signals.~ 
$^1$\protect\cite{cfw94},~
$^2$\protect\cite{tml93},~
$^3$\protect\cite{lnl+95},~ 
$^4$\protect\cite{cnt93},~
$^5$\protect\cite{bhl+94},~ 
$^6$\protect\cite{lzc95},~
$^7$\protect\cite{dt91},~
$^8$\protect\cite{nt95},~
$^9$\protect\cite{nll+95},~ 
$^{10}$\protect\cite{cam95a},~ 
$^{11}$\protect\cite{bbm+95},~ 
$^{12}$\protect\cite{arz95}.
} 
\label{t:pbdot}
\end{table}   


\begin{thebibliography}{}

\bibitem[Applegate \& Shaham 1994]{as94}
Applegate, J.~H. and Shaham, J. 1994, \apj, 436, 312.

\bibitem[Arzoumanian 1995]{arz95}
Arzoumanian, Z. 1995.
\newblock PhD thesis, Princeton University.

\bibitem[Arzoumanian, Fruchter \& Taylor 1994]{aft94}
Arzoumanian, Z., Fruchter, A.~S., and Taylor, J.~H. 1994, \apj, 426, L85.

\bibitem[Bailes { et al.}  1994]{bhl+94}
Bailes, M. { et al.}  1994, \apjlett, 425, L41.

\bibitem[Bell { et al.}  1995]{bbm+95}
Bell, J.~F., Bailes, M., Manchester, R.~N., Weisberg, J.~M., and Lyne, A.~G.
  1995, \apj, 440, L81.

\bibitem[Blandford, Romani \& Applegate 1987]{bra87}
Blandford, R.~D., Romani, R.~W., and Applegate, J.~H. 1987, \mnras, 225, 51{P}.

\bibitem[Camilo 1995]{cam95a}
Camilo, F. 1995.
\newblock PhD thesis, Princeton University.

\bibitem[Camilo, Foster \& Wolszczan 1994]{cfw94}
Camilo, F., Foster, R.~S., and Wolszczan, A. 1994, \apj, 437, L39.

\bibitem[Camilo, Nice \& Taylor 1993]{cnt93}
Camilo, F., Nice, D.~J., and Taylor, J.~H. 1993, \apjlett, 412, L37.

\bibitem[Camilo, Thorsett \& Kulkarni 1994]{ctk94}
Camilo, F., Thorsett, S.~E., and Kulkarni, S.~R. 1994, \apj, 421, L15.

\bibitem[Damour \& Taylor 1991]{dt91}
Damour, T. and Taylor, J.~H. 1991, \apj, 366, 501.

\bibitem[Flynn \& Fuchs 1994]{ff94}
Flynn, C. and Fuchs, B. 1994, \mnras, 270, 471.

\bibitem[Frail \& Weisberg 1990]{fw90}
Frail, D.~A. and Weisberg, J.~M. 1990, \aj, 100, 743.

\bibitem[Gwinn { et al.}  1986]{gtwr86}
Gwinn, C.~R., Taylor, J.~H., Weisberg, J.~M., and Rawley, L.~A. 1986, \aj, 91,
  338.

\bibitem[Iben \& Tutukov 1986]{it86}
Iben, I. and Tutukov, A.~V. 1986, \apj, 311, 742.

\bibitem[Lorimer { et al.}  1995]{lnl+95}
Lorimer, D.~R. { et al.}  1995, \apj, 439, 933.

\bibitem[Lundgren, Zepka \& Cordes 1995]{lzc95}
Lundgren, S.~C., Zepka, A.~F., and Cordes, J.~M. 1995, \apj.
\newblock In press.

\bibitem[Nicastro { et al.}  1995]{nll+95}
Nicastro, L., Lyne, A.~G., Lorimer, D.~R., Harrison, P.~A., Bailes, M., and
  Skidmore, B.~D. 1995, \mnras, 273, L68.

\bibitem[Nice \& Taylor 1995]{nt95}
Nice, D.~J. and Taylor, J.~H. 1995, \apj, 441, 429.

\bibitem[Ryba \& Taylor 1991]{rt91a}
Ryba, M.~F. and Taylor, J.~H. 1991, \apj, 371, 739.

\bibitem[Taylor \& Cordes 1993]{tc93}
Taylor, J.~H. and Cordes, J.~M. 1993, \apj, 411, 674.

\bibitem[Taylor, Manchester \& Lyne 1993]{tml93}
Taylor, J.~H., Manchester, R.~N., and Lyne, A.~G. 1993, \apjsupp, 88, 529.

\bibitem[Will 1993]{wil93}
Will, C.~M. 1993, { Theory and Experiment in Gravitational Physics},
  (Cambridge: Cambridge University Press).

\bibitem[Wolszczan 1994]{wol94}
Wolszczan, A. 1994, 264, 538.

\end{thebibliography}
\end{document}